\documentclass[prr,twocolumn,floatfix,superscriptaddress]{revtex4}

\usepackage{lscape}
\usepackage{rotating}
\usepackage{epstopdf}
\usepackage{graphicx}
\usepackage{natbib}
\usepackage{epstopdf}
\usepackage{amsmath}
\usepackage{amssymb}
\usepackage{mathbbol}
\usepackage{xcolor}
\usepackage{lipsum}
\usepackage{amssymb}
\usepackage{mwe}
\usepackage{lipsum} 
\usepackage{siunitx}
\usepackage[normalem]{ulem}

\usepackage{empheq}
\usepackage{afterpage}
\usepackage[colorlinks=true,citecolor=blue]{hyperref}
\hypersetup{colorlinks=true,citecolor=blue,linkcolor=blue,urlcolor=blue}
\usepackage{cleveref}

\newcommand{\e}{\mathrm{e}}

\newcommand{\beq}{\begin{equation}}
\newcommand{\eeq}{\end{equation} \smallskip}
\newcommand{\beqy}{\begin{eqnarray}}
\newcommand{\eeqy}{\end{eqnarray} \smallskip}

\newcommand{\bit}{\begin{itemize}}
\newcommand{\eit}{\end{itemize}}
\newcommand{\bmat}{\begin{pmatrix}}
\newcommand{\emat}{\end{pmatrix}}

\begin{document}

\title{Non-Hermitian topological end-mode lasing in polariton systems}

\author{P. Comaron}
\affiliation{Institute of Physics, Polish Academy of Sciences, Al. Lotnik\'ow 32/46, 02-668 Warsaw, Poland}

\author{V. Shahnazaryan}
\affiliation{Institute of Physics, Polish Academy of Sciences, Al. Lotnik\'ow 32/46, 02-668 Warsaw, Poland}
\affiliation{ITMO University, St. Petersburg 197101, Russia}

\author{W. Brzezicki}
\affiliation{International Research Centre MagTop, Institute of Physics, Polish Academy of Sciences, Al. Lotnik\'ow 32/46, 02-668 Warsaw, Poland}

\author{T. Hyart}
\affiliation{International Research Centre MagTop, Institute of Physics, Polish Academy of Sciences, Al. Lotnik\'ow 32/46, 02-668 Warsaw, Poland}

\author{M. Matuszewski}
\affiliation{Institute of Physics, Polish Academy of Sciences, Al. Lotnik\'ow 32/46, 02-668 Warsaw, Poland}

\date{\today}

\begin{abstract}
We predict the existence of non-Hermitian topologically protected end  states in a one-dimensional exciton-polariton condensate lattice, where topological transitions are driven by the laser pump pattern.
We show that
the number of end states can be described by a Chern number and a topological invariant based on the Wilson loop.
We find that such transitions  arise due to {\it enforced exceptional points} which can be predicted directly from the bulk Bloch wave functions. This allows us to establish a new type of bulk-boundary correspondence for non-Hermitian systems and to compute the phase diagram of an open chain analytically. 
Finally, we demonstrate topological lasing of a single end-mode in a realistic model of a microcavity lattice. 
\end{abstract}

\maketitle

Systematic analysis of symmetries and topological invariants has revealed a rich variety of topological materials which possess robust boundary states stemming from the topology of the  bulk bands  \cite{HasanReview, Chiu2016}. The next endeavour is to extend these concepts to the non-Hermitian (NH) systems~\cite{Esaki2011, Schomerus13, Lee2016, Menke2017,Ke2017,Xu2017,Ni2018,ZhouL2018,LiKitaev2018, MartinezAlvarez2018, Gong2018, Zhou19PRB, Kaw18}. While many interesting properties not present in their Hermitian counterparts were predicted to exist~\cite{Leykam2017,Yao2018,YaoBand2018,Kunst2018,Shen2018, MATORRES18, Yokozimo2019, Brzezicki19}, NH topological states were realized experimentally only in a handful of systems~\cite{Zeuner2015,Zhan2017,Xiao2017,Weimann2017, stJean2017, Bahari2017, Parto2018, Bandres2018, Zhao2018,Zhou2018}. The vast majority of the theoretical proposals require violation of reciprocity, which is difficult to implement experimentally in both condensed matter and optical systems. Recently it was demonstrated that topologically protected end modes can arise in lattice systems where non-Hermiticity can be expressed as complex-valued diagonal on-site terms \cite{Takata2018,Lieu2018,Yoshida2019}.  A possible realization consists of a lattice of coupled optical waveguides~\cite{Parto2018,Bandres2018} where the non-trivial topology originates from the periodic modulation of the gain-loss balance. 

In this Rapid Communication we study NH physics in a light-matter hybrid quantum quasiparticle system  \cite{CarusottoReview} and demonstrate the feasibility of NH topological end states and end mode lasing. The non-trivial topology is achieved through spatial modulation of external incoherent pump intensity in a one-dimensional lattice of  homogeneously coupled microcavity exciton-polaritons. These are natural candidate systems for realizing topological phases \cite{Karzig2015,Nalitov2015}, but in existing proposals \cite{Bardyn2015,Bleu2016,Banerjee2018,Janot_TopologicalHallCavity,Malpuech_TopologicalGapSoliton,Savenko_TIMagneticDots,Malpuech_ZTopologicalInsulator,Liew_FloquetTopologicalPolaritons,Shelykh_TopologicalMetamaterialsRings,Kartashov_2DTopologicalPolaritonLaser,Kartashov_BistableTI,Downing_TopologicalPhasesCavityWaveguide,Liew_ChiralBogoliubons,Liew_SpontaneousChiralEdgeStates} and realizations~\cite{stJean2017,Bloch_TopologicalInvariantsQuasicrystals,Amo_OrbitalEdgeStates,klembt2018,Whittaker2018} topological order arises from the band-topology in a Hermitian Hamiltonian, so it is not related to the open dissipative NH nature of the system. We show that the presence of excitonic component opens up a possibility to realize a polariton system with competing Hermitian and non-Hermitian effects, so that unique topological phases possessing various number of topological end states can be observed. In particular limits of the model the number of end states is described by a Chern number \cite{Brzezicki19} and a topological invariant based on the eigenvalues of the Wilson loop  \cite{Alex14}, but by continuity the end states exist beyond these special cases. We find that the end modes can disappear or appear at transitions where the energy gap between the bulk bands and the end states closes without closing of the bulk energy gap. 
Such transitions can occur in open NH systems due to high order exceptional points \cite{MATORRES18} leading to the NH skin effect \cite{Yao2018,YaoBand2018,Kunst2018,Thomale2019,Yokozimo2019}.
However, we find that skin effect is not present in our model because the Hamiltonian satisfies a specific type of NH time-reversal symmetry \cite{Kaw18}. We discover that the transitions occurring in an infinite open chain arise because specific bulk Bloch wave functions are incompatible with the boundary conditions leading to {\it enforced exceptional points} in the spectrum. This way we establish a new type of bulk-boundary correspondence for NH Hamiltonians, because the changes in the number of end states can be directly predicted from the bulk Bloch wave functions. This powerful new tool  allows us to compute the topological phase diagram of an infinite open system analytically.

Moreover, we demonstrate the possibility of lasing of end modes for realistic physical parameters corresponding to a lattice of polariton micropillars. By appropriate choice of pumping intensity pattern, we achieve the situation when only the selected end state is amplified, which results in topological single mode lasing.

\begin{figure}[h!]
	\centering
		    \includegraphics[width=\linewidth]{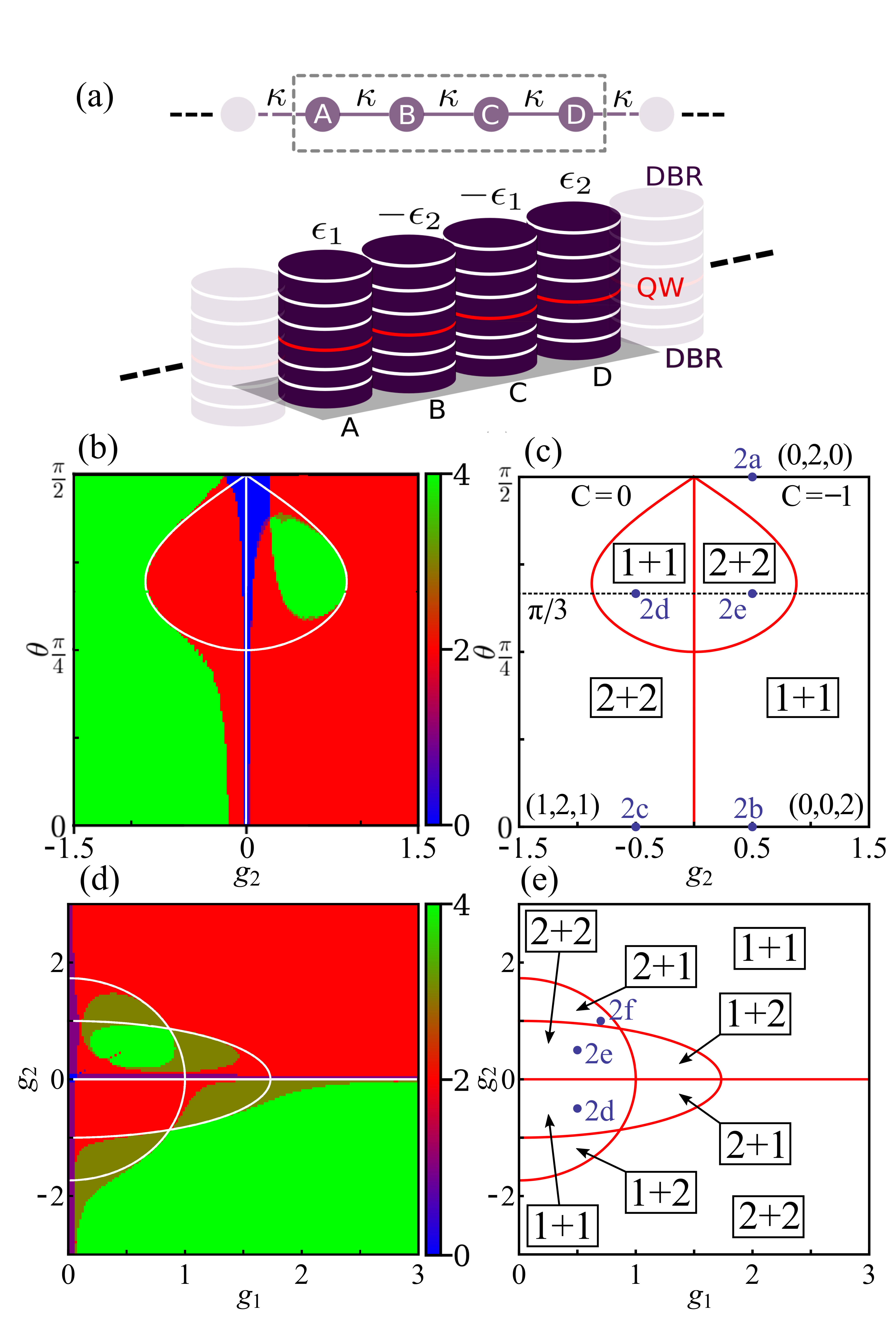}
    	\caption{
    	(a) The non-Hermitian four-site unit cell lattice of microcavity pillars with embedded quantum wells.
    	(b) Topological phase diagram 	for the  symmetric system $g_1 = |g_2|$   obtained by counting the end states as a function of pumping modulation amplitudes $g_{1,2}$ and non-Hermiticity parameter $\theta$. The length of the chain in the numerical calculation is $L=4\times 100$.
    	(c) Topological phase diagram for an infinite open chain with $g_1 = |g_2|$ obtained by combining the information from Chern number $C$, Wilson loop invariant and enforced exceptional points. The phases with $M_L$ ($M_R$) localized modes  in the left (right) end of the chain are denoted as $[M_L+M_R]$.  In the special cases where the bulk bands and the end states can be ordered based on the real part of the energies we use a notation $(m_1,m_2,m_3)$, where $m_i$ is the number of end states in the $i$th energy gap. 
    	(d),(e) Numerically and analytically computed phase diagrams as a function of $g_1$ and $g_2$ for $\theta = \pi/3$. 
    	The small disagreement between the phase diagrams arises due to the finite size effects. All energies are in units of $\kappa$.  
    	}
    	\label{fig:phasediagrams}
\end{figure}

We consider a one-dimensional (1D) lattice of coupled micropillars as depicted in~Fig.~\ref{fig:phasediagrams}(a). Each micropillar contains a quantum well and is assumed to host a tightly bound exciton-polariton mode~\cite{kavokin2017microcavities}. We start with the system of discrete mean-field Gross-Pitaevskii equations~\cite{Wouters2007,Stepnicki_TightBinding}
\begin{align}
    \label{eq:GPE}
  i \hbar \dot{\psi}_n&= -\kappa\sum_{\langle nn \rangle} \psi_m + \left[g_c|\psi_n|^2+g_R n_n^R+i  {\hbar} \frac{Rn_n^R-\gamma_c}{2}\right]\psi_n,\nonumber \\
  \dot{n}_n^R&=P_n - \left(\gamma_R + R |\psi_n|^2\right)n_n^R,
\end{align}
where $\psi_n(t)$ is the condensate amplitude in the $n$-th lattice cell, the $\langle nn\rangle$ sum runs over nearest neighbors, $n_n^R(t)$ is the density of exciton reservoir in the $n$-th cell, $P_n$ is the external nonresonant pumping rate, $\gamma_c$ and $\gamma_R$ are the decay rates of the condensate and the reservoir, respectively, $g_c$ and $g_R$ are the corresponding interaction constants, and $R$ is the  rate of scattering from the reservoir to the condensate. We assume that the polariton interactions within the condensate represented by the term $g_c|\psi_n|^2$ are negligible in comparison with the reservoir-condensate interaction $g_Rn_n^R$, which is a good approximation in most experiments where nonresonant pumping is used.

While the coupling coefficients $\kappa$ are assumed to be the same for each pair of neighboring micropillars, the external pumping $P_n$ is modulated spatially with the four-site periodicity. 
Therefore, the pumping gives rise to complex on-site potential [Fig.~\ref{fig:phasediagrams}(a)]. 
Within the adiabatic approximation \cite{supp,Bobrovska2015Adiabatic}, this leads to an effective NH tight-binding Hamiltonian 
 \begin{equation}
	\hat{H}_k (g_1,g_2) = 
	\begin{pmatrix}
	g_1 e^{i \theta} & \kappa & 0 & \kappa e^{- i k}  \\
	\kappa &  - g_2 e^{i \theta} & \kappa & 0   \\
	0  & \kappa  & - g_1 e^{i \theta} & \kappa   \\
	\kappa e^{i k} & 0 & \kappa & g_2 e^{i \theta} 	
	\end{pmatrix}.
	\label{eq:Ham_k}
\end{equation}
 Here $k$ is the Bloch wave number in the units of inverse intercell distance, $\epsilon_{1,2}=g_{1,2}\e^{i \theta}$ are the complex on-site potentials
which are given by $\epsilon_n=\left(n_n^R - \gamma_c/R\right)(g_R+i  {\hbar} R/2)$.
The real part of $\epsilon_{1,2}$ stands for the interaction-induced on-site potential, whereas the imaginary part corresponds to the balance between  gain and loss. In the linear approximation of negligibly small $|\psi_n|^2$, which we assume in most of this work, the reservoir occupation $n_n^R$ in the above equation is defined by the local pumping rate and takes the form $n_n^R=n_{n0}^R = P_n/\gamma_R$, and we have chosen the pump rates $P_n$ so that the onsite potentials in every second lattice sites are related to each other. The real amplitudes $g_n$ can be positive or negative depending on the local pumping intensity~\cite{supp}.  For realistic repulsive interactions between polaritons one has $0 <\theta< \pi/2$.  When complex on-site potentials are absent ($\epsilon_{1,2}=0$), the lattice is trivial since all the coupling coefficients $\kappa$ are equal.
In typical polariton micropillar lattices $\kappa \sim 0.1\,$meV.

\begin{figure}[t]
    \centering
	\includegraphics[width=\columnwidth]{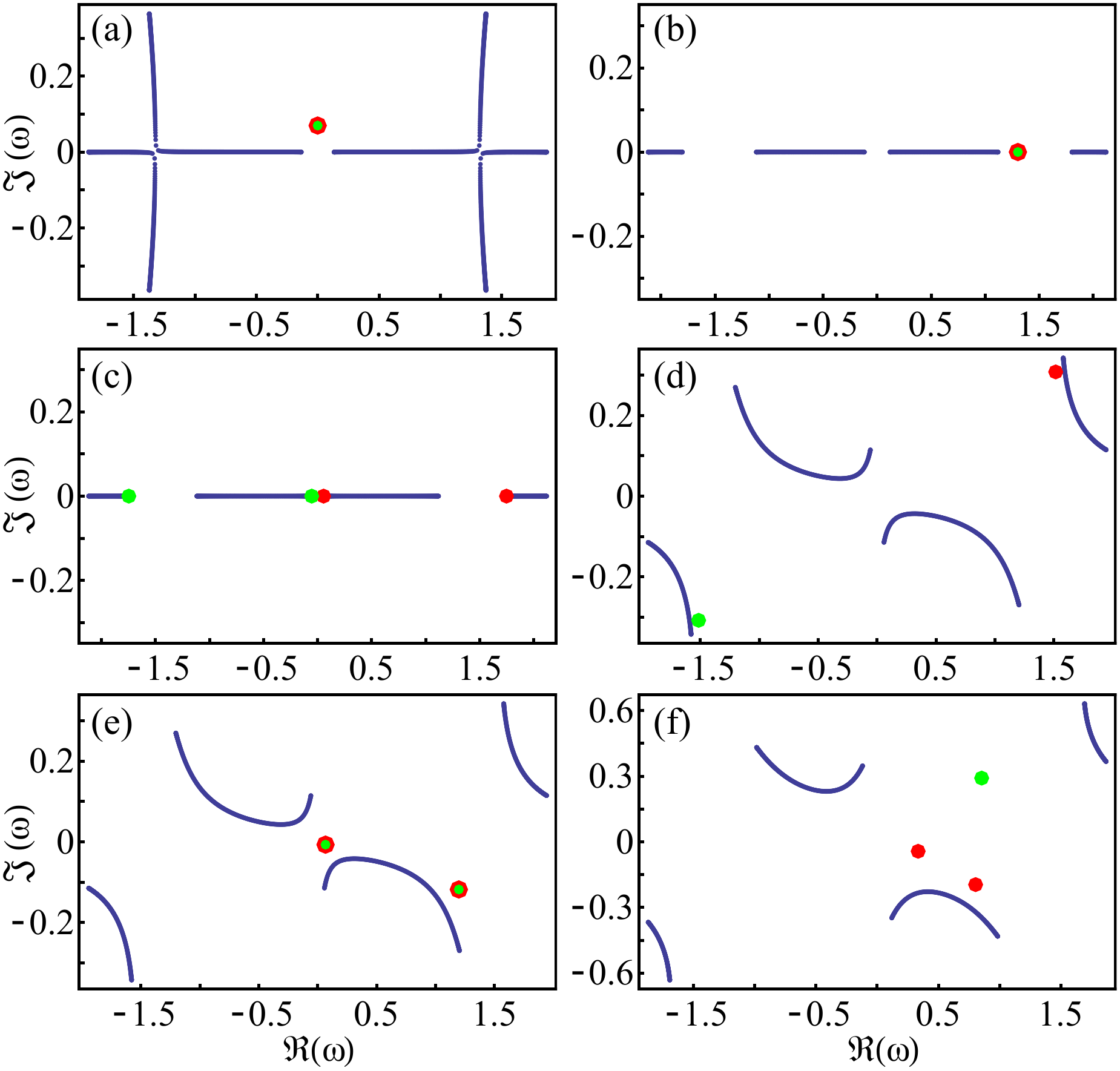}
	\caption{Complex energy spectra of an open chain with $L=1000$ for the topological phases shown in Figs.~\ref{fig:phasediagrams}(c),(e). The red/green dots denote end states at left/right end
of the chain. The parameters are
(a) $g_1=g_2=0.5$, $\theta=\pi/2$,
(b) $g_1=g_2=0.5$, $\theta=0$,
(c) $-g_1=g_2=-0.5$, $\theta=0$,
(d) $-g_1=g_2=-0.5$, $\theta=\pi/3$,
(e) $g_1=g_2=0.5$, $\theta=\pi/3$,
(f) $g_1=0.7$, $g_2=1.0$, $\theta=\pi/3$.
All energies are in units of $\kappa$.}
	\label{fig:edge_states}
\end{figure}

We solved the Hamiltonian (\ref{eq:Ham_k}) numerically with open boundary conditions and calculated  the number of end states  as a function of various parameters of the model~\cite{supp}.  The results are shown in Figs.~\ref{fig:phasediagrams} and \ref{fig:edge_states}. We denote the phases with $[M_L+M_R]$, where $M_L$ ($M_R$) is the number of modes localized in the left (right) end.  In the special cases where we can order the bulk bands and the end states based on the real part of the energies we also use a notation $(m_1,m_2,m_3)$, where $m_i$ is the number of end states in the energy gap between $i$th and $i+1$th bands.

In the extreme NH limit $\theta=\pi/2$ the system obeys NH chiral symmetry ${\cal S}\hat{H}_{k}(g_1, g_2){\cal S}=-\hat{H}^{\dagger}_{k}(g_1, g_2)$ with ${\cal S}=\mathbb{1}\otimes\sigma_{z}$. In this case the number of end states with zero real part of the energy is described by a Chern number $C$ \cite{Brzezicki19, supp}. For $g_2>0$ and $g_1=|g_2|$ we obtain $C=-1$ which means that there should be one state with zero real part of the energy at each end of the chain in agreement with our numerical calculations [Figs.~\ref{fig:phasediagrams}(c) and \ref{fig:edge_states}(a)]. Thus, this limiting case of the model belongs to the [1+1] phase with the number of end states in each energy gap given by $(0,2,0)$.  In this limit the system
exhibits topological states solely due to the non-Hermiticity  \cite{Takata2018}.  

If  $\theta=0$, $g_1=|g_2|$ and $g_2>0$ the system is Hermitian and obeys an inversion symmetry $\mathcal{I} \hat{H}_k (g,g) \mathcal{I}^{-1} =\hat{H}_{-k} (g,g)$, where $\mathcal{I}=\sigma_x\otimes\sigma_{x}$. Therefore, a topological invariant can be defined based on a Wilson loop invariant or quantized Zak phases  \cite{Alex14, supp}. We find that both approaches yield identical predictions \cite{supp}. Namely, we find that the first two energy gaps are trivial whereas the third energy gap is nontrivial [Fig.~\ref{fig:edge_states}(b)]. Thus, the number of end states in each energy gap given by $(0,0,2)$ [Fig.~\ref{fig:phasediagrams}(c)].

If $\theta=0$, $g_1=|g_2|$ and $g_2<0$ the bulk Hamiltonian obeys inversion symmetry but the unit cell is not compatible with the inversion symmetry. 
This obscures the bulk-boundary correspondence but non-rigorous reasoning based on Wilson loop invariants indicates that the model belongs to [2+2] phase with the number of end states $(1,2,1)$ \cite{supp}, in agreement with numerical calculations [Fig.~\ref{fig:edge_states}(c)].

 Although in a realistic description of exciton-polaritons one has $0<\theta < \pi/2$, the limits considered above are usueful because 
end states can disappear or appear only when their energy approaches bulk bands. Therefore, we  obtain the overall phase diagram shown in Fig.~\ref{fig:phasediagrams}(b),(c). There is an energy gap closing between bands if $g_1=g_2=0$, so this phase transition is also easy to understand.

We find that if $\pi/4 < \theta < \pi/2$  there exists additional transitions where the end modes disappear or appear  without closing of the bulk  gap [Fig.~\ref{fig:phasediagrams}(b),(c)]. Such  transitions are known to occur due to the NH skin effect \cite{Yao2018,YaoBand2018,Kunst2018,Yokozimo2019}, but the skin effect is absent in our model because of NH time-reversal symmetry $\hat{H}^*_{-k} (g_1,g_2)=\hat{H}^\dag_{k} (g_1,g_2)$ \cite{Kaw18}. Instead, we notice that the eigenstates of the continuum bands in an infinite open chain  can be constructed from standing waves of bulk Bloch wave functions. This construction follows the one suggested in Ref.~\cite{Yokozimo2019}, but we discover that if the first or last component of a bulk Bloch wave function vanishes it is impossible to satisfy boundary conditions. Therefore, states should disappear from continuum bands at parameter values where such incompatibility occurs. 
A connection between boundary conditions and exceptional points was also pointed out in Ref.~\cite{Foa_Torres_2019} but the relation to the bulk wave functions (i.e. bulk-boundary correspondence) was not discovered in this earlier work.
This can happen in NH systems with the help of exceptional points which transform into new end states. In the following we demonstrate that with the help of this reasoning we can compute the topological phase diagram of an infinite open system analytically.

To find the {\it enforced exceptional points} in an infinite open chain we need to calculate when the product $Q_k$ of the first and the last components of all eigenvectors  vanishes. 
In the case $g_1=|g_2|$ we find {by a straightforward calculation} that 
\[
Q_k \propto f_kf_{-k}, \  f_{k}=4g_2^{2}e^{i(k+2\theta)}+\kappa^{2} (1+e^{ik}-e^{2ik}-e^{3ik}).
\]
Therefore, $Q_k$ vanishes along the contour
\begin{equation}
g_2= \kappa \frac{\sin k}{\sqrt{2\left|\sin\frac{k}{2}\right|}}, \ \theta=\frac{\pi+\left|k\right|}{4}, \ k\in [-\pi,\pi). \label{transition1}
\end{equation}
The vanishing $Q_k$ leads to appearance of two exceptional points because of inversion (chiral-inversion) symmetry if $g_2>0$ ($g_2<0$). Therefore, the number of end states changes by 2 at these transition lines. By computing the end state spectrum for $g_2>0$ ($g_2<0$) we indeed find that [1+1] ([2+2]) phase transforms into [2+2] ([1+1]) phase when $g_2$ and $\theta$ are tuned across the transition line defined by Eq.~(\ref{transition1}) [Fig.~\ref{fig:phasediagrams}(b),(c) and Fig.~\ref{fig:edge_states}(d),(e)]. In addition to the prediction of the phase boundaries, we can predict the momentum and the band where the exceptional point leads to appearance or disappearance of end states. We checked that numerical calculations are in agreement with these analytical considerations{ , see \cite{supp}.}

Similarly we can find the zeros of  $Q_k$ in the  $g_1$-$g_2$ plane for fixed $\theta$. The explicit
expression for $Q_k$ is more complicated but we arrive to a simple result that $Q_k$ vanishes along
the four elliptic contours
\begin{equation}
g_{1}=\pm \kappa  \cos\frac{k}{2}, \ g_{2}= \pm \kappa \tan\theta\sin\frac{k}{2}, \ k\in [-\pi,\pi)  \nonumber \\
\end{equation}
and
\begin{equation}
g_{1}=\pm \kappa  \tan\theta\sin\frac{k}{2}, \ g_{2} = \mp \kappa \cos\frac{k}{2}, \ k\in [-\pi,\pi).  \nonumber \\
\end{equation}
In the general case $g_1 \ne |g_2|$ the vanishing of  $Q_k$ leads to a single exceptional point so that the number of end states changes by 1 at these transition lines. Thus, we obtain exotic phases [1+2] and [2+1], with different number of topological states at the two ends of the chain.  Our numerical calculations are in agreement with these analytic considerations [Fig.~\ref{fig:phasediagrams}(d),(e),(f)]. Thus, our results demonstrate that the transitions occurring in an infinite open chain can be predicted from Bloch wave functions, establishing a new type of bulk-boundary correspondence for non-Hermitian systems. 
 
We now consider the possibility of lasing of these end states. We expect that single-mode lasing can be realized if all modes decay in time except a selected end mode, which is amplified by the medium. This requires an adjustment of the spectra presented in Fig.~\ref{fig:edge_states}. As demonstrated in Supplementary Information~\cite{supp}, the following set of coupled equations describes evolution in the weakly nonlinear regime
\begin{equation}
          i \hbar \dot{\psi_n} = 
    - \kappa \sum_{\langle nn \rangle} \psi_m +
    {\epsilon}_n \psi_n 
    - \Gamma_n\left( 1 + i \tan \theta \right)|\psi_n|^2 \psi _n,
    \label{eq:dynamics}  
\end{equation}
where now we include additional loss $\gamma$ in each node, i.~e.~$\epsilon_{\rm A}=\epsilon_1-i {\hbar}\gamma$,  $\epsilon_{\rm B}=-\epsilon_2-i {\hbar}\gamma$,  $\epsilon_{\rm C}=-\epsilon_1-i {\hbar}\gamma$, and $\epsilon_{\rm D}=\epsilon_2-i {\hbar}\gamma$. This results in the shift of the imaginary part of the excitation spectrum by $- {\hbar}\gamma$. For a careful choice of $\gamma$,
we can stabilize all the eigenstates except one. Physically, it corresponds to homogeneous reduction of pumping across the lattice. The reduction of pumping affects also the real part of the potential, but this can be eliminated by a rotating frame for the condensate amplitudes~\cite{supp}. The parameter $\Gamma_n = P_n R g_R/\gamma_R^2$ describes the nonlinear interactions with the reservoir.

\begin{figure}[t!] 
	\centering
	\includegraphics[width=\linewidth]{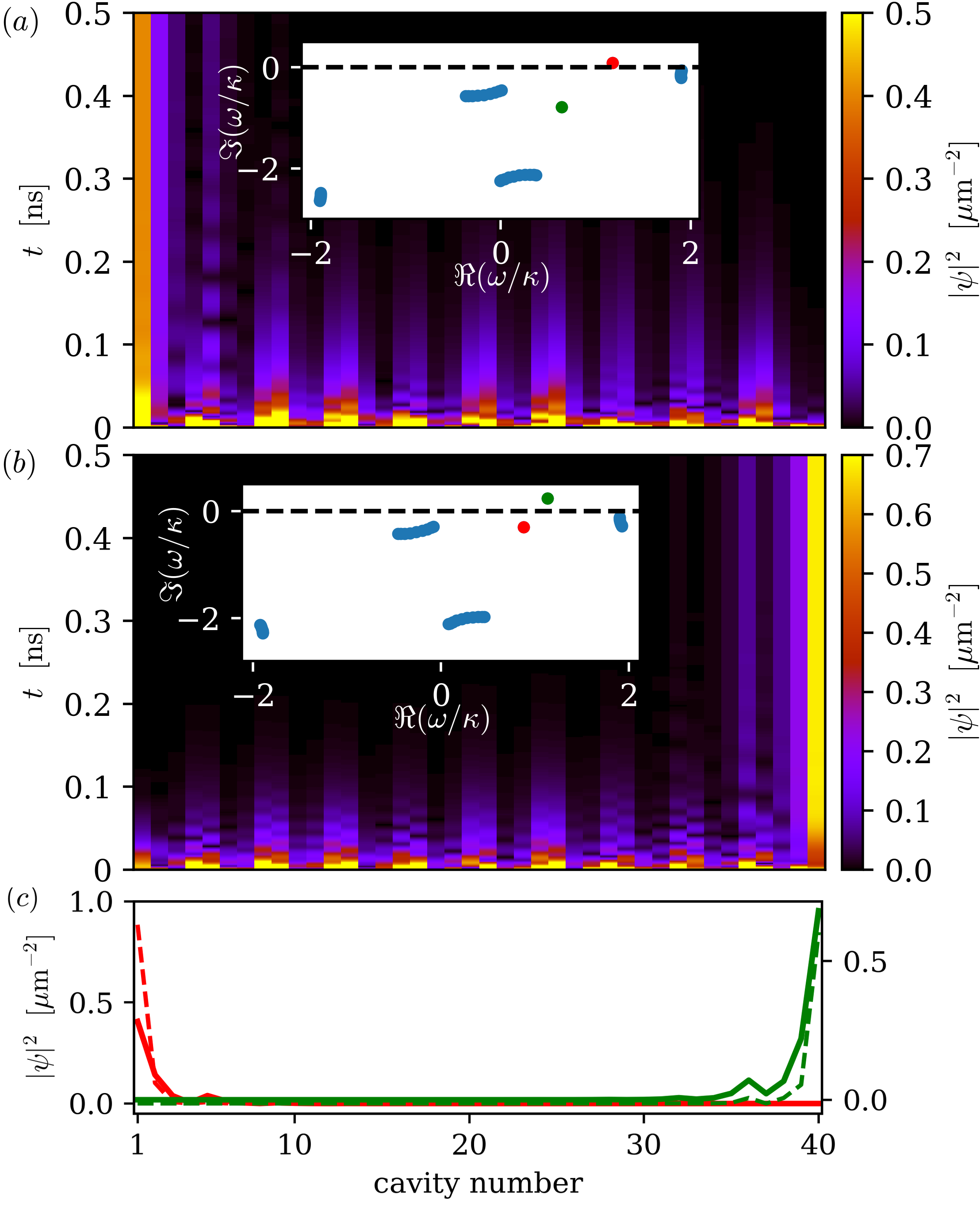}
	\caption{
		The emergence of end mode lasing in the nonlinear model~(\ref{eq:dynamics}). 
		The color scale shows the evolution of 
		the field density in each cavity for physical parameters 
		(a) $g_1 =0.2 \mathrm{meV}$, $g_2 = 0.1 \mathrm{meV}$  and $  {\hbar} \gamma=0.135 \mathrm{meV}$; 
		(b) $g_1 =0.14 \mathrm{meV}$, $g_2 = 0.14 \mathrm{meV}$, $  {\hbar}\gamma=0.12 \mathrm{meV}$. In the rightmost pillar we apply an extra pump so that there the value of $g_2$ is modified to $g_2 = 0.196 \mathrm{meV}$.
		Insets in (a) and (b) show the corresponding spectra of open chain in the linear limit as in Fig.~\ref{fig:edge_states}.
		(c) Steady state density profiles after long evolution time for the cases shown in panel (a) (solid red line) and (b) (solid green line) are compared to the corresponding end-states densities (dashed lines) calculated from the eigenvalues in the linear limit.
		Other parameters are $\kappa = 0.1 \ \mathrm{meV}$,  {$\gamma_c  = 1.52 \ \mathrm{ps^{-1}}$, $\gamma_R=0.075 \ \mathrm{ps^{-1}}$, $R = 6.9 \times 10^{-3} \ \mathrm{ps^{-1} \mu m^2}$}, $g_R =1.35 \times 10^{-3} \ \mathrm{meV \mu m^2}$, $\theta=\pi/3$ and $L=10$.
		%
	}
	\label{fig:edgeLasing}
\end{figure}

In Fig.~\ref{fig:edgeLasing} we show examples of the evolution of the intensity distributions obtained by solving Eq.~\eqref{eq:dynamics} in a $L = 4\times 10$ unit-cell configuration with a random initial amplitude. 
After initial evolution, the polariton density saturates at a steady-state distribution, which approximately corresponds to the end mode in the linear spectrum, see~Fig.~\ref{fig:edgeLasing}(c).

In the symmetric case with $g_1 = |g_2|$ end states are always below bulk states in the imaginary part of the spectrum. To establish single-end mode lasing, one has to apply additional pump to one of the end pillars~\cite{stJean2017}. This leads to the increase of the imaginary part of the corresponding end state, relative to all other states. The resulting imaginary part of the spectrum is shown in the inset of \ref{fig:edgeLasing}(b). Here the values of on-site effective potentials $g=\pm 0.14 \mathrm{meV}$ can be created with external pumps $P_+ \approx {  16.45 \mu\mathrm{m^{-2}ps^{-1}} }$, and $P_- \approx {  8.45 \mu\mathrm{m^{-2}ps^{-1}} }$. In addition, we apply extra pump to the rightmost pillar with $\delta P\approx {  6.58 \mu\mathrm{m^{-2}ps^{-1}} }$.

In conclusion, we predicted a new type of bulk-boundary correspondence for non-Hermitian systems, based on enforced exceptional points. These topological modes can be realized in realistic polariton micropillar lattices. Our general model applies to any non-Hermitian system where competition between gain, loss and interactions exists, such as in cold atom systems. 

\

We acknowledge support from the National Science Center, Poland grant No.~2016/22/E/ST3/00045 and grant No.~2017/25/Z/ST3/03032 under QuantERA program. VS acknowledges support from the mega-grant No. 14.Y26.31.0015 of the Ministry of Education and Science of the Russian Federation.
The research was partially supported by the Foundation for Polish Science through the IRA Programme co-financed by EU within SG OP.

%




	
	\pagebreak
	
	\cleardoublepage
	
	\widetext
	\begin{center}
		\textbf{\large Supplementary Material for: 	\\ Non-Hermitian topological end-mode lasing in polariton systems}
	\end{center}
	
	\vspace{8mm}
	
	\setcounter{equation}{0}
	\setcounter{figure}{0}
	\setcounter{table}{0}
	\setcounter{page}{1}
	\renewcommand{\theequation}{S\arabic{equation}}
	\renewcommand{\thefigure}{S\arabic{figure}}
	\renewcommand{\bibnumfmt}[1]{[S#1]}
	\renewcommand{\citenumfont}[1]{S#1}
	
	\twocolumngrid
	
	\

	{In this Supplementary Material we present more details related to the  numerical and theoretical procedure presented in the main paper.}
	
	\
	
	\paragraph*{{Derivation from the polariton model.-}}
	
	The system of coupled micropillars with polariton resonance is described by  Eq.~(1) of the main text. In the approximation of negligibly small polariton interactions, the first of the two equations can be expressed as
	\beq
	i \hbar \dot{\psi}_n= -\kappa\sum_{\langle nn \rangle} \psi_m + \left[\left(g_R+\frac{i  {\hbar}}{2}R\right)n_n^R-\frac{i {\hbar}}{2}\gamma_c\right]\psi_n. \label{S1}
	\eeq
	The prefactor of last term is equivalent to the on-site effective energy of the tight-binding Hamiltonian (2) in the main text. 
	One has
	\begin{align} 
		u_n&=u_n^{Re}+iu_n^{Im}= g_R n_n^R + \frac{i  {\hbar}}{2}\left(R n_n^R - \gamma_c\right) =\nonumber\\
		&= \left(n_n^R - \frac{\gamma_c}{R}\right)\left(g_R+\frac{i {\hbar}}{2}R\right) + \frac{g_R\gamma_c}{R}.\label{epsilons}
	\end{align}
	The last term in the above is a constant real energy shift, which can be removed by introducing a rotating frame for condensate amplitudes, $\psi_n\rightarrow \psi_n{\rm e}^{-i(g_R\gamma_c/R)t}$. In result, we obtain the effective complex potentials in the form $\epsilon_n=g_n{\rm e}^{i\theta}$, where
	\begin{align}
		\epsilon_n^{Re} = \left(n_n^R - \frac{\gamma_c}{R}\right)g_R, \nonumber\\
		\epsilon_n^{Im} = \left(n_n^R - \frac{\gamma_c}{R}\right)\frac{ {\hbar} R}{2}, 
	\end{align}
	with $\tan \theta = \frac{ {\hbar} R}{2g_R}$. It is clear that while $g_{1,2}$ can be either positive or negative, the parameter $\theta$ must fulfill $\tan \theta >0$. 
	In the linear approximation of negligibly small $|\psi_n|^2$, $n_n^R$ in the above equation is simply proportional to the local pumping rate and takes the form
	\beq
	n_n^R=n_{n0}^R = \frac{P_n}{\gamma_R}.
	\eeq
	
	It should be noted that the  Eq.~(\ref{S1}) can be obtained from the effective tight-binding Hamiltonian
	\begin{eqnarray}
		\hat{H} =  \sum_n  \left( 
		\epsilon_1 \hat{a}^\dagger_n \hat{a}_n  
		- \epsilon_2 \hat{b}^\dagger_n \hat{b}_n 
		- \epsilon_1 \hat{c}^\dagger_n \hat{c}_n 
		+ \epsilon_2 \hat{d}^\dagger_n \hat{d}_n 
		\right) +  \nonumber \\
		\kappa \sum_n  \left( 
		\hat{b}^\dagger_n \hat{a}_n + \hat{c}^\dagger_n \hat{b}_n + \hat{d}^\dagger_n \hat{c}_n + \hat{a}^\dagger_{n+1} \hat{d}_n 
		+ H.c. \right) ,
		\label{eq:Hamiltonian}
	\end{eqnarray}
	where $\hat{a}_n$, $\hat{b}_n$, $\hat{c}_n$ and $\hat{d}_n$ ($\hat{a}^\dagger_n$, $\hat{b}^\dagger_n$, $\hat{c}^\dagger_n$ and $\hat{d}^\dagger_n$) denote annihilation (creation) operators for the sub-lattices $A$, $B$, $C$ and $D$ in the $n$th lattice cell of our structure, 
	$1 \leq n \leq L$. Here $L$ is the number of unit cells,  $4L$ is the total number of sites, and $\epsilon_i$ are the complex effective potentials. By going to the momentum space we arrive to the Eq.~(2) in the main text.
	To go beyond the linear approximation, one may employ the adiabatic approximation~\cite{Bobrovska2015Adiabatic} where
	\beq
	n_n^R = \frac{P_n}{\gamma_R+R|\psi_n|^2}\approx \frac{P_n}{\gamma_R} - \frac{P_n R}{\gamma_R^2}|\psi_n|^2 + O(|\psi_n|^4),
	\eeq
	in the first order approximation with respect to the condensate density.
	
	\
	
	\paragraph*{{Numerical procedure for end mode detection.-}}
	
	In the main text we discuss presence of topologically protected end modes in the case of  open boundary conditions [Fig.~1(b),(d) and Fig.~2 in main text]. For each values of the parameters $(g_1,g_2,\theta)$, we have first calculated eigenvalues and eigenstates by solving the eigenvalue problem for the Hamiltonian (2) in the main text, and then classified the eigenstates to bulk and end modes by evaluating the ratio between the integral of the intensity in the most far left (and most far right) unit cell with the total eigenmodes norm. For bulk (end) states this ratio goes to zero (non-zero) in the limit $L\to \infty$, but for finite $L$ we identify the states as end states if the ratio is above a chosen threshold. We have used $L=100$ and threshold ratios $0.04$ and $0.15$ in Fig.~1(b) and  Fig.~1(d), respectively.
	
	\
	
	{ 
		\paragraph*{{Enforced exceptional points and end states.-}}
		
		\begin{figure}[t]
			\includegraphics[width=\linewidth]{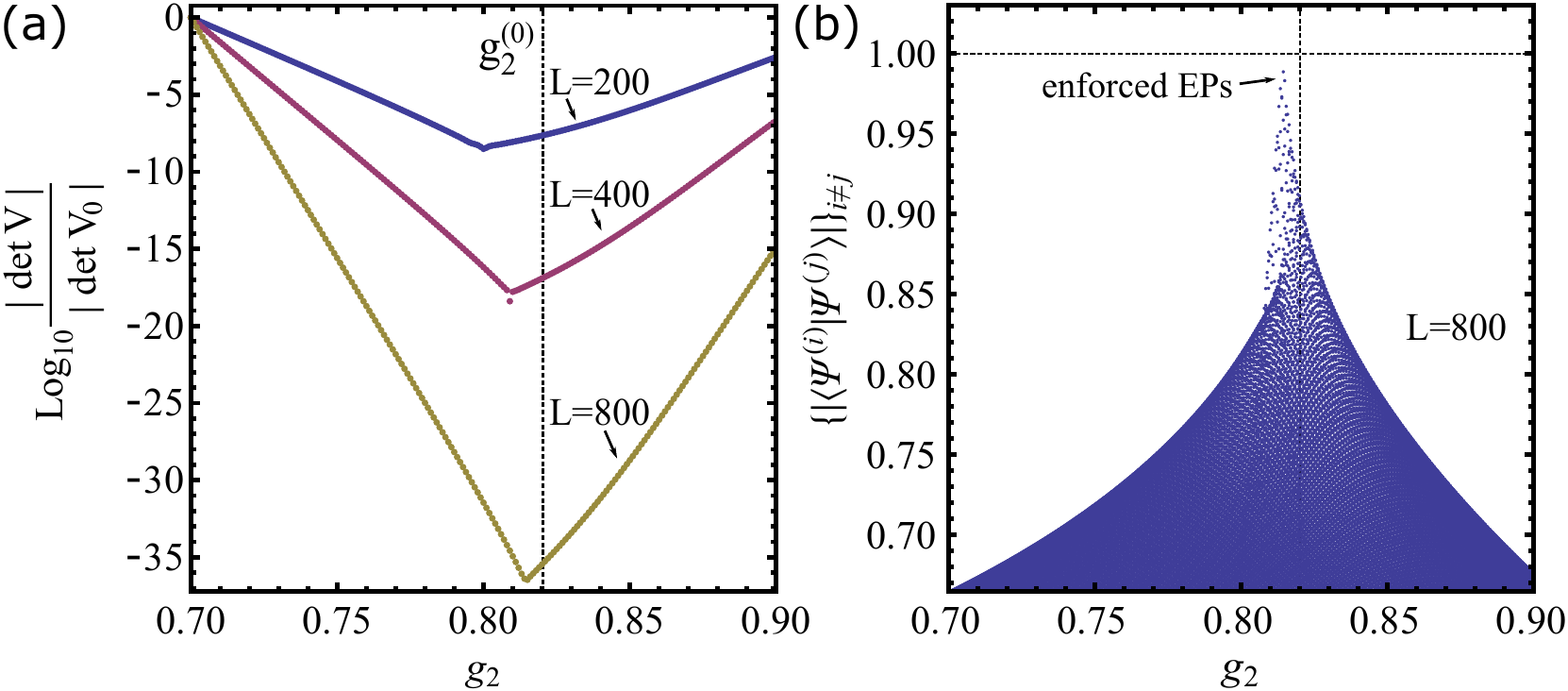}
			\caption{{ 
					(a) Evolution of the determinant of the normalized eigenvectors matrix $V$ of the Hamiltonian as a function of $g_2=g_1$ for $\theta =1.2$  		
					and  $L=200,400,800$. The determinant is normalized by a reference value $\det|V(g_2=0.7)|\equiv \det|V_0|$. (b) Evolution of the ovelaps between the normalized eigenvectors of the Hamiltonian $|\langle\psi^{(i)}|\psi^{(j)}\rangle |$ (where $i\not =j$) as a function of $g_2=g_1$ for $\theta =1.2$  		
					and  $L=800$. The maximal value $1$ means the appearance of an exceptional point where two eigenvectors become parallel to each other. These numerical results indicate that in the limit $L \to \infty$ the exceptional point appears at the critical point $g_2 = g_2^{(0)}$ where the first or last component of the Bloch wave function vanishes. 
				}}
				\label{fig:EP}
			\end{figure}
			
			In this section we discuss our numerical evidence that whenever the first or last component of the bulk Bloch wave function vanishes, we always find an enforced exceptional point developing in the thermodynamic limit $(L\to \infty)$ in an infinite open chain, and this leads to appearance/disappearance of end states. We now consider one of the phase transition points shown in Fig.~1(c) in detail. 
			In Fig.~\ref{fig:EP} we show the behavior of the determinant of the eigenvectors matrix of the Hamiltonian and the overlaps of all pairs of different eigenvectors as a function of 
			$g_2$ across the phase transition point. In Fig.~\ref{fig:EP}(a) we show that the determinant goes to zero exponentially with increasing $L$  at the critical point $g_2 \approx g_2^{(0)}$ where the first or last component of the Bloch wave function vanishes, indicating the appearance of the enforced exceptional point. The appearance of the exceptional points is also confirmed in Fig.~\ref{fig:EP}(b) where we show that the maximum overlap between the eigenvector approaches $1$ as $g_2 \to g_2^{(0)}$ and $L\to \infty$.  We have checked that the two states that give the largest overlaps are the two end states which become parallel to two different bulk states. This directly demonstrates that the end states emerge from the enforced exceptional points. Further numerical evidence of the connection of the enforced exceptional points and the appearance/disappearance of end states is shown in Fig.~1 in the main text. The numerically calculated number of end states always changes at the transition points where the bulk Bloch wave function component vanishes.}
		
		\
		
		\paragraph*{{Chern number.-}}
		
		In the case $\theta=\pi/2$ the Hamiltonian (2) in the main text satisfies a NH chiral symmetry
		\begin{equation}
			{\cal S}\hat{H}_{k}(g_1, g_2){\cal S}=-\hat{H}^{\dagger}_{k}(g_1, g_2),\label{eq:chir}
		\end{equation}
		with ${\cal S}=\mathbb{1}\otimes\sigma_{z}$. This means that we can define  a bulk topological invariant which describes the number of end modes~\cite{Brzezicki19}. Namely, we can obtain an
		effective two-dimensional Hermitian Hamiltonian as
		\begin{equation}
			H^{{\rm eff}}\left(k,\eta\right)={\cal S}\left(\eta-i\hat{H}_{k}(g_1, g_2)\right) \label{eq:heffk}
		\end{equation}
		and in the case of open boundary conditions the zero-energy end states of $H^{{\rm eff}}\left(k,\eta\right)$ correspond to end states of $\hat{H}_{k}(g_1, g_2)$ with zero real part of energy and imaginary part equal to $\eta$ \cite{Brzezicki19}. Because $H^{{\rm eff}}\left(k,\eta\right)$ is defined in a two-dimensional $\left(k,\eta\right)$ space and its Chern number $C$ is quantized if it is gapped,  $C\ne 0$ at half-filling implies that the Hamiltonian $\hat{H}_{k}(g_1, g_2)$ supports $|C|$ end states with zero real part of the energy \cite{Brzezicki19}. The Chern
		number $C$ for $H^{{\rm eff}}\left(k,\eta\right)$ can
		be obtained using Kubo formula 
		\begin{equation}
			C=\frac{1}{2 \pi}\int_{-\infty}^{+\infty}d\eta\int_{0}^{2\pi}dk\Omega_{k,\eta},
		\end{equation}
		where the Berry curvature $\Omega_{k,\eta}$ is given by
		\begin{equation}
			\Omega_{k,\eta}=\!\!\sum_{{n\le n_{F}\atop m>n_{F}}}\!\!\mathfrak{Im}\frac{2\left\langle \psi_{k,\eta}^{n}\right|\!\partial_{k}H^{{\rm eff}}\!\left|\psi_{k,\eta}^{m}\right\rangle \!\left\langle \psi_{k,\eta}^{m}\right|\!\partial_{\eta}H^{{\rm eff}}\!\left|\psi_{k,\eta}^{n}\right\rangle }{\left(E_{k,\eta}^{(n)}-E_{k,\eta}^{(m)}\right)^{2}}.
		\end{equation}
		Here $\left|\psi_{k,\eta}^{p}\right\rangle $ are eigenstates of $H^{{\rm eff}}\left(k,\eta\right)$ (sorted in ascending order of eigenenergy) and $n_{F}$ is the number of occupied bands.
		
		\
		
		\paragraph*{{Wilson loop invariant.-}} If $g_1=|g_2|$ the bulk Hamiltonian obeys an inversion symmetry $\mathcal{I}$. We focus here on the case $g_2>0$.  In this case $\mathcal{I} \hat{H}_k (g,g) \mathcal{I}^{-1} =\hat{H}_{-k} (g,g)$, where $\mathcal{I}=\sigma_x\otimes\sigma_{x}$. Therefore, we can define a  topological invariant $N_{(-1)}$ using the ideas of Ref.~\cite{Alex14}. The most
		general definition makes use of Wilson loop operator given by the
		matrix elements
		\begin{equation}
			{\cal W}_{ij}=\left\langle \psi^{(i)}_{-\pi}\right|{\cal P}_{-\pi+\delta k}{\cal P}_{-\pi+2\delta k}\dots{\cal P}_{\pi}\left|\psi^{(j)}_{-\pi}\right\rangle ,
		\end{equation}
		where $i,j=1,\dots,n_{F}$ label the occupied bands $\left|\psi^{(i)}_{k}\right\rangle $
		of ${\cal H}_{k}$,  and ${\cal P}_{k}$ is the projector
		on occupied bands for a given $k$. The step size $\delta k$
		has to be chosen sufficiently small. For non-orthogonal eigenvectors
		projector ${\cal P}_{k}$ can be easily constructed after
		orthogonalizing the set of occupied eigenvectors $\left\{ \left|\psi_{i}\left(k\right)\right\rangle \right\} _{i=1}^{n_{F}}$
		to obtain new vectors $\left\{ \left|\tilde{\psi}_{i}\left(k\right)\right\rangle \right\} _{i=1}^{n_{F}}$
		that span the same subspace so that ${\cal P}_{k}=\sum_{i=1}^{n_{F}}\left|\tilde{\psi}_{i}\left(k\right)\right\rangle \left\langle \tilde{\psi}_{i}\left(k\right)\right|$.
		According to \cite{Alex14} in the Hermitian case and in the
		presence of inversion symmetry the spectrum of ${\cal W}$ consists
		of eigenvalues that are either $\pm1$ or pairs of complex conjugate
		numbers $\lambda$ and $\lambda^{\star}$. The topological invariant
		$N_{(-1)}$ is then given by the number of $-1$ eigenvalues of ${\cal W}$.
		We have checked that the spectra
		of ${\cal W}$ hold the same properties in the NH case and hence we can define the
		$N_{(-1)}$ invariant in an analogous way. An important simplification
		coming from \cite{Alex14} is that the invariant can be
		expressed as
		\begin{equation}
			N_{(-1)}=\left|n_{(-1)}(0)-n_{(-1)}(\pi)\right|,
		\end{equation}
		where $n_{(-1)}(0)$ and $n_{(-1)}(\pi)$ are the numbers of occupied
		states at $k=0$ and $k=\pi$ having inversion eigenvalue $-1$ (${\cal I}$
		commutes with ${\cal H}_{k}$ at these points). We have checked that
		in our case this formula also gives identical results as the Wilson loop
		approach. 
		
		We note that in order to use these invariants in description of non-Hermitian systems one needs to be able to keep track which bands are occupied as a function of the parameters of the model. 
		In the main text we use these invariants only in the Hermitian case where the bands can be ordered based on the energies. This way we can define a topological invariant for each energy gap by assuming that all bands below the gap are occupied when calculating $N_{(-1)}$. By collecting the invariants calculated for the different gaps we obtain a Wilson loop invariant $\{N_{(-1), 1}, N_{(-1),2}, N_{(-1),3} \}$.
		
		The approach can be easily generalized to all situations where the bands can be ordered according to the real (or imaginary) parts of the energies. In a more general situation, one needs to be able to follow the spectral flow of the bands as a function of the parameters of the model.
		
		\
		
		\paragraph*{{Zak phases.-}}
		The gauge-invariant prescription for a Zak
		phase of band $i$ reads $\nu_{i}=\exp(i\varphi_{i})$ with
		\begin{align}
			\varphi_{i} & = \arg[ \left\langle \psi^{(i)}_{-\pi}\right|\left.\psi^{(i)}_{-\pi+\delta k}\right\rangle \left\langle \psi^{(i)}_{-\pi+\delta k}\right|\left.\psi^{(i)}_{-\pi+2\delta k}\right\rangle \dots \nonumber\\
			&  \dots \,  \left. \left\langle \psi^{(i)}_{\pi-\delta k}\right|\left.\psi^{(i)}_{-\pi}\right\rangle \right].
		\end{align}
		Now, assuming inversion symmetry and  $\delta k\to0$ one can show that 
		\begin{equation}
			\nu_{i}=\left\langle \psi^{(i)}_{-\pi}\right|{\cal I}\left|\psi^{(i)}_{-\pi}\right\rangle \left\langle \psi^{(i)}_{0}\right|{\cal I}\left|\psi^{(i)}_{0}\right\rangle ,
		\end{equation}
		proving the quantization of Zak phase and providing simplification
		with respect to original formula. 
		The Zak phases of the four bands seem to predict correctly the (non)triviality
		of a given energy gap in the sense that when the bands can be ordered based on the real part of the energy we have
		the following correspondence between Wilson loop invariants of the energy gaps
		and Zak phases of the bands
		\[
		\begin{array}{ccc}
		\{0,0,1\} & \leftrightarrow & \left(+1,+1,-1,-1\right),\\
		\{0,1,1\} & \leftrightarrow & \left(+1,-1,+1,-1\right),\\
		\{1,1,0\} & \leftrightarrow & \left(-1,+1,-1,+1\right).
		\end{array}
		\]
		In terms of the Zak phases the energy gap is topologically nontrivial if the product of the Zak phases of the occupied bands is $-1$.
		
		\
		
\begin{figure}[t]
			\includegraphics[width=\linewidth]{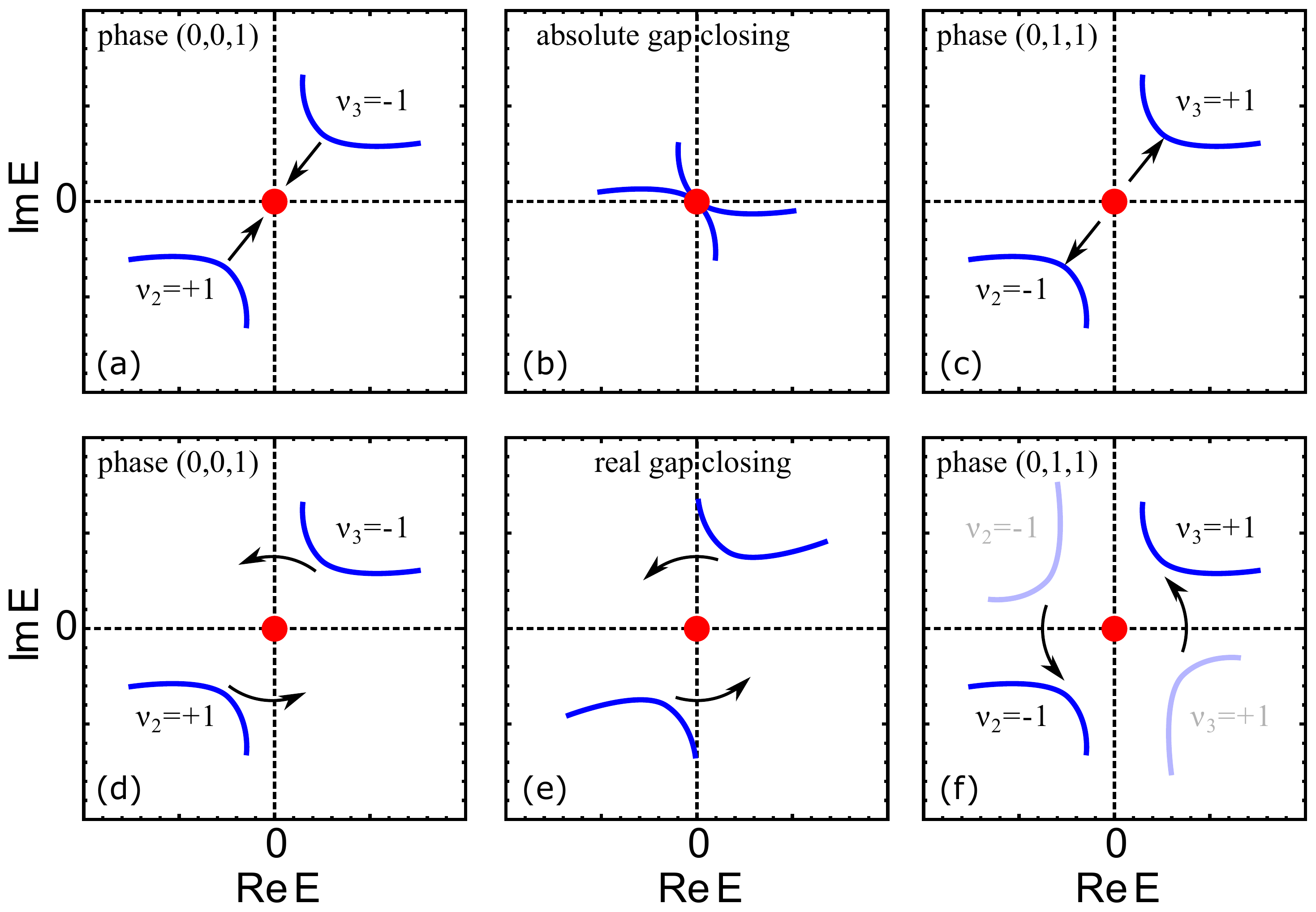}
			\caption{
				Schematic view of changing bulk topological phase $\left(0,0,1\right)$
				to $\left(0,1,1\right)$ by point gap closing (a-c) or by  line
				gap closing (d-f). Only bands $i=2,3$  their Zak phases $\nu_{i}$ are shown.
			}
			\label{fig:bands}
		\end{figure}

		\paragraph*{{Point and line gap closing.-}} If one does not keep track of the spectral flow of the occupied bands in the calculation of the Zak phases there is an important difference between non-Hermitian and Hermitian systems. Namely, in  Hermitian systems the Zak phases can only change when two bands touch each other, but in non-Hermitian systems 
		the Zak phases can change in two different ways:  ($i$) bands touch in the complex-energy plane (point
		gap closing) and exchange their invariants  or ($ii$)
		bands are interchanged in the complex plane without touching but with closing of
		a line energy gap (Fig.~\ref{fig:bands}). If one follows the spectral flow of the bands as a function of the parameters of the model and keeps the same bands occupied the Zak phases can only change in a point gap closing.
		
		\
		
		\paragraph*{{Dependence of the number of end modes on Wilson loop invariant in the Hermitian limit.-}} The important question is of course how the number of end modes $(m_1, m_2, m_3)$ obtained in the different gaps are related to the Wilson loop topological invariant $\{N_{(-1), 1}, N_{(-1),2}, N_{(-1),3} \}$ in the Hermitian limit $\theta=0$. 
		
		If $g_1=|g_2|$ and $g_2>0$ we find that $m_i=2 N_{(-1), i}$ ($i=1,2,3$) because each nontrivial Wilson loop invariant gives rise to topological state at both ends of the chain. In the case of our model we obtain $N_{(-1), 1}=N_{(-1),2}=0$ and $N_{(-1),3}=1$ so that the number of end modes in each energy gap are $(0,0,2)$.
		
		However, the situation is slightly more complicated if $g_1=|g_2|$ and $g_2<0$. In this case the bulk Hamiltonian obeys inversion symmetry but the unit cell is not compatible with the inversion symmetry. This means that in the case of open boundary conditions one needs to include additionally half of a unit cell in the end of the chain in order to make the Hamiltonian inversion symmetric. This obscures the bulk-boundary correspondence because one of the ends comes from a bulk Hamiltonian where parameters are replaced as $g_i \to -g_i$. Numerically, we find that for $\theta=0$,  $g_1=|g_2|$ and $g_2<0$ the Wilson loop invariants are $\{1, 1, 0 \}$ but for opposite signs of the couplings $g_i$ they are $\{0, 1, 1 \}$. If we assume that each end of the chain obeys one set of Wilson loop invariants we arrive to the number of end modes $(1,2,1)$ in agreement with our numerical calculations. The end modes need to be symmetrically distributed in the gaps because the Hamiltonian obeys a chiral-inversion symmetry $\mathcal{C} \hat{H}_k (g,g) \mathcal{C}^{-1} =-\hat{H}_{-k} (g,g)$ with $\mathcal{C}=\sigma_x\otimes\sigma_{y}$, which makes the spectrum particle-hole symmetric.
		
		\
		
		\paragraph*{Global Berry phase.-}
		The global Berry phase is defined as~\cite{LiangGlobalBerryPhase}
		\begin{equation}\label{eq:GlobalBerry}
			W=-\frac{i}{4\pi} \sum\limits_{l=1}^4 \int\limits_{-2\pi}^{2\pi} \mathrm{d} \varphi \langle \psi_{l,L} | \partial_\varphi |\psi_{l,R} \rangle,     
		\end{equation}
		where $|\psi_{l,R[L]} \rangle$ denotes the normalized $l$-th right [left] eigenstate of the NH bulk Hamiltonian [Eq.~(2) in the main text]. The global Berry phase is quantized to be an integer but it depends on the gauge choice, and therefore it cannot describe any observables. The discontinuities of the global Berry phase in a fixed gauge can correspond to topological phase transitions in NH systems~\cite{LiangGlobalBerryPhase}, but the utility of this property is limited by the fact that the discontinuities may also appear due to the breakdown of the gauge convention. Despite these problems, the global Berry phase has been utilized in many papers~\cite{Takata2018,Bandres2018,Yao2018}, so we discuss it also in the context of our work. 
		
\begin{figure}[t]
			\centering
			\includegraphics[width=\linewidth]{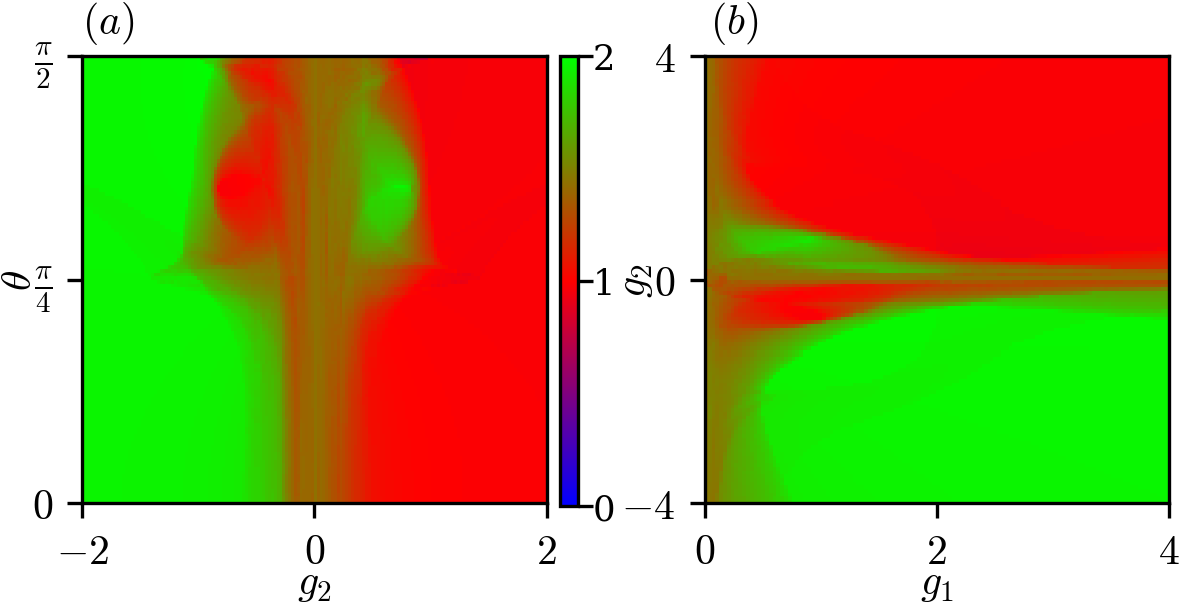}
			\caption{
				The global Berry phase calculated in phase spaces $(g_2, \theta)$, for (a) $g_1 = |g_2|$ and  (b) $(g_1, g_2)$ with $\theta = \pi/3$.
			}
			\label{fig:SM_Fig1}
		\end{figure}

		We use a fixed gauge where the first component of the wave function is real and positive. 
		With this (optimally chosen) gauge convention, the numerical value of the global Berry phase~(\ref{eq:GlobalBerry}) in the parameter space $(g_2, \theta)$ with $g_1=|g_2|$ leads to a good agreement  with the numerically calculated number of end states (see Fig. \ref{fig:SM_Fig1} (a)). The value of global Berry phase $W=2$ corresponds to two topological states at each end, and $W=1$ to one pair of end states. Interestingly, the global Berry phase is not only able to identify the points in the phase diagram where the gap between different bands closes but it changes also at the topological phase boundaries corresponding to the enforced exceptional points. This happens accidentally because our gauge convention breaks down at the values of the parameters where the first component of the wave function vanishes. In the vicinity of the phase boundaries the numerically calculated $W$ takes intermediate non-quantized values due to the limited accuracy of the numerical integration.  
		
		We further go beyond the symmetric case and study the situation when $g_1 \neq g_2$. The results for $W$ are shown in  Fig \ref{fig:SM_Fig1}(b).  In this case the results of global Berry phase calculations are not in  agreement with numerical determination of number of end states. Therefore, we conclude that even with optimally chosen gauge convention the global Berry phase gives only partial description for the number of topological end states. 
		

\end{document}